\documentclass[a4paper,11pt]{article}
\usepackage{pos}

\title{Update of kaon semileptonic form factor using $N_f=2+1$ PACS10 configurations}
\ShortTitle{Update of kaon semileptonic form factor using $N_f=2+1$ PACS10 configurations}

\author*[a,b]{Takeshi Yamazaki}

\author[c]{Ken-ichi~Ishikawa}

\author[b]{Naruhito~Ishizuka}

\author[b]{Yoshinobu~Kuramashi}

\author[d]{Yusuke~Namekawa}

\author[b]{Yusuke~Taniguchi}

\author[a]{Naoya~Ukita}

\affiliation[]{\normalsize{\bf \sffamily \hspace{50mm} (PACS Collaboration)}}

\affiliation[a]{Institute of Pure and Applied Sciences, University of Tsukuba, Tsukuba, Ibaraki 305-8571, Japan}

\affiliation[b]{Center for Computational Sciences, University of Tsukuba, Tsukuba, Ibaraki 305-8577, Japan}

\affiliation[c]{Core of Research for the Energetic Universe, Graduate School of Advanced Science and Engineering, Hiroshima University, Higashi-Hiroshima, 739-8526, Japan}

\affiliation[d]{Education and Research Center for Artificial Intelligence and Data Innovation, Hiroshima University, Higashi-Hiroshima 739-8521, Japan}

\emailAdd{yamazaki@het.ph.tsukuba.ac.jp}

\abstract{
We calculate the form factors for the kaon semileptonic decay process
using the PACS10 configurations, whose physical volume is more than 
(10 fm)$^4$ very close to the physical point. The configurations 
were generated with the Iwasaki gauge action and $N_f=2+1$ stout-smeared 
nonperturbatively $O(a)$-improved Wilson quark action at the three lattice 
spacings, 0.085, 0.063, and 0.041 fm. We present updated results for the form
factors, and discuss their continuum extrapolations, momentum transfer
interpolation, and short chiral extrapolation to tune the simulated 
pion and kaon masses to the physical ones. From the results with various
analyses, the systematic error of the form factor at the zero momentum
transfer is estimated. The value of $|V_{us}|$ is 
determined using our result, and is compared with those using the previous 
calculations and also those determined through the kaon leptonic decay 
process.}

\FullConference{The 41st International Symposium on Lattice Field Theory (LATTICE2024)\\
 28 July - 3 August 2024\\
Liverpool, UK\\}

\begin{document}
\maketitle

\section{Introduction}

The Cabibbo-Kobayashi-Maskawa (CKM) matrix elements in 
the first row give 2.2 $\sigma$ violation from 
the CKM unitarity~\cite{ParticleDataGroup:2022pth}.
It could suggest the existence of physics beyond the standard model.
The value of the violation is essentially evaluated by $|V_{ud}|$ and $|V_{us}|$.
The kaon leptonic ($K_{\ell 2}$) and semileptonic ($K_{\ell 3}$)
decay processes are adopted to obtain the value of $|V_{us}|$,
while it is reported that these values have a little 
tension~\cite{ParticleDataGroup:2022pth}.
To understand the violation of the CKM unitarity,
it is important to investigate the difference between the two determinations
of $|V_{us}|$.

For the $K_{\ell 2}$ decay determination, $|V_{us}|$ is basically given by
the decay constant ratio $F_K/F_\pi$, 
which is well measured in the lattice QCD calculation.
On the other hand, the $K_{\ell 3}$ form factor at the zero momentum transfer 
$f_+(0)$ is necessary to obtain $|V_{us}|$ for the $K_{\ell 3}$ determination.
Since this calculation is more complicated than that of the decay constants,
various studies have been carried out to
obtain precise values of the $K_{\ell 3}$ form factor~\cite{Dawson:2006qc,Boyle:2007qe,Lubicz:2009ht,Bazavov:2012cd,Boyle:2013gsa,Bazavov:2013maa,Boyle:2015hfa,Carrasco:2016kpy,Aoki:2017spo,Bazavov:2018kjg}.

Using the $N_f = 2+1$ PACS10 configurations generated on huge volumes 
very close to the physical point, we also measured 
the $K_{\ell 3}$ form factor~\cite{PACS:2019hxd,Ishikawa:2022ulx}
at the two larger lattice spacings.
Although the accuracy of the result is comparable with
the previous calculations, 
our result has a large systematic error coming from
an estimate of the lattice spacing effect.
To reduce the systematic error, we performed the calculation
at the third lattice spacing~\cite{Yamazaki:2023swq}.
In this report, we present updated results in our calculation from
the last lattice conference, although all the results at the
smallest lattice spacing are still preliminary.
To obtain $f_+(0)$ from data of the $K_{\ell 3}$ form factors in finite 
momentum transfers at three lattice spacings,
we perform a simultaneous fit for the momentum transfer interpolation,
continuum extrapolation, and also short chiral extrapolation to
the physical point.
A systematic error of $f_+(0)$ for the continuum extrapolations 
is estimated by using various analyses.
We also determine the value of $|V_{us}|$ using our preliminary result.

\section{Simulation parameters}

The PACS10 configurations were generated using 
the Iwasaki gauge action~\cite{Iwasaki:2011jk}
and a non-perturbative $O(a)$-improved Wilson quark action 
with the six-stout-smeared link~\cite{Morningstar:2003gk}.
The simulation parameters of the PACS10 configurations at the three lattice
spacings are tabulated in Table~\ref{tab:sim_param}.

\begin{table}[!b]
\caption{
Simulation parameters of the PACS10 configurations 
at the three lattice spacings.
The bare coupling ($\beta$), lattice size ($L^3\cdot T$),
physical spatial extent ($L$), lattice spacing ($a$),
the number of the configurations ($N_{\rm conf}$), 
pion and kaon masses ($m_\pi$, $m_K$), and the range of the timeslice
separation ($t_{\rm sep}$) between the source and sink operators 
in the three-point function are tabulated.
  \label{tab:sim_param}
}
\begin{center}
\begin{tabular}{ccccccccc}\hline\hline
Label & $\beta$ & $L^3\cdot T$ & $L$[fm] & $a$[fm] &
$N_{\rm conf}$ & $m_\pi$[MeV] & $m_K$[MeV] & $t_{\rm sep}$[fm] \\\hline
PACS10/L128 &
2.20 & 256$^4$ & 10.5 & 0.041 & 20 & 142 & 514 & 3.4-3.9 \\
PACS10/L160 &
2.00 & 160$^4$ & 10.1 & 0.063 & 20 & 138 & 505 & 2.3-4.1 \\
PACS10/L256 &
1.82 & 128$^4$ & 10.9 & 0.085 & 20 & 135 & 497 & 3.1-4.0 \\\hline\hline
\end{tabular}
\end{center}
\end{table}

The same quark action is employed in the measurements 
for the two- and three-point functions for the $K_{\ell 3}$ form factors.
The details of the calculation method in our study are described in 
Ref.~\cite{Ishikawa:2022ulx}.
Using the correlation functions, the matrix elements for 
the $K_{\ell 3}$ form factors are extracted, which are written by
the two form factors $f_+(q^2)$ and $f_-(q^2)$ as,
\begin{eqnarray}
\langle \pi (\vec{p}_{\pi}) \left | V_{\mu} \right | K(\vec{p}_{K}) \rangle = ({p}_{K}+{p}_{\pi})_{\mu}f_{+}(q^2)+ ({p}_{K}-{p}_{\pi})_{\mu}f_{-}(q^2),
\label{eq:def_matrix_element}
\end{eqnarray}
where $V_\mu$ is the weak vector current and $q^2$ is 
the momentum transfer squared.
Another form factor $f_0(q^2)$ is defined with the two form factors as,
\begin{eqnarray}
f_{0}(q^2) = f_{+}(q^2) + \frac{-q^2}{{m^2_{K}}-{m^2_{\pi}}}f_{-}(q^2).
\label{eq:f0}
\end{eqnarray}
In this study, we calculate $f_+(q^2)$ and $f_0(q^2)$
with six and seven different $q^2$, respectively, around $q^2 = 0$
at each lattice spacing.

For the calculation of the two- and three-point functions,
we adopt the random source operator spread in the spin, color, and
spatial spaces proposed in Ref.~\cite{Boyle:2008yd}.
The exponentially smeared quark operator with the random source is
also employed for the calculations of the correlation functions
at the two smaller lattice spacings.
The several time separations, $t_{\rm sep}$,
between the source and sink operators 
in the three-point function are utilized to investigate excited state
contamination of the matrix elements.
The range of $t_{\rm sep}$ at each lattice spacing 
is shown in Table~\ref{tab:sim_param}.
The form factors are evaluated with a combined analysis using
the correlation functions with the two source operators
and the different $t_{\rm sep}$.

The periodic boundary condition is imposed in the spatial directions
in the calculation of the correlation functions,
while in the temporal direction the periodic and anti-periodic
boundary conditions are employed.
The average of the two-point correlation functions with the different boundary
conditions makes the periodicity of the temporal direction effectively doubled.
A similar average of the three-point functions suppresses
the wrapping-around effect~\cite{Kakazu:2017fhv}
in the small momentum region~\cite{PACS:2019hxd,Ishikawa:2022ulx}.

Not only the local weak vector current but also the conserved vector one 
is employed to calculate the three-point functions
to investigate finite lattice spacing effects in the form factors.
The renormalization factor of the local vector current is determined from
$Z_V = \sqrt{Z_V^\pi Z_V^K}$ where
\begin{equation}
Z_V^H = \frac{1}{F^{\rm bare}_H(0)}
\label{eq:zv}
\end{equation}
for $H = \pi, K$ with $F^{\rm bare}_H(0)$ being 
the bare electromagnetic form factor
evaluated using the local vector current at $q^2 = 0$.

\section{Preliminary results}

In this section our preliminary results for the $K_{\ell 3}$ form factors and
also the value of $|V_{us}|$ determined from our results
are presented.

\subsection{$K_{\ell 3}$ form factors}

\begin{figure}[!t]
 \centering
 \includegraphics*[scale=0.41]{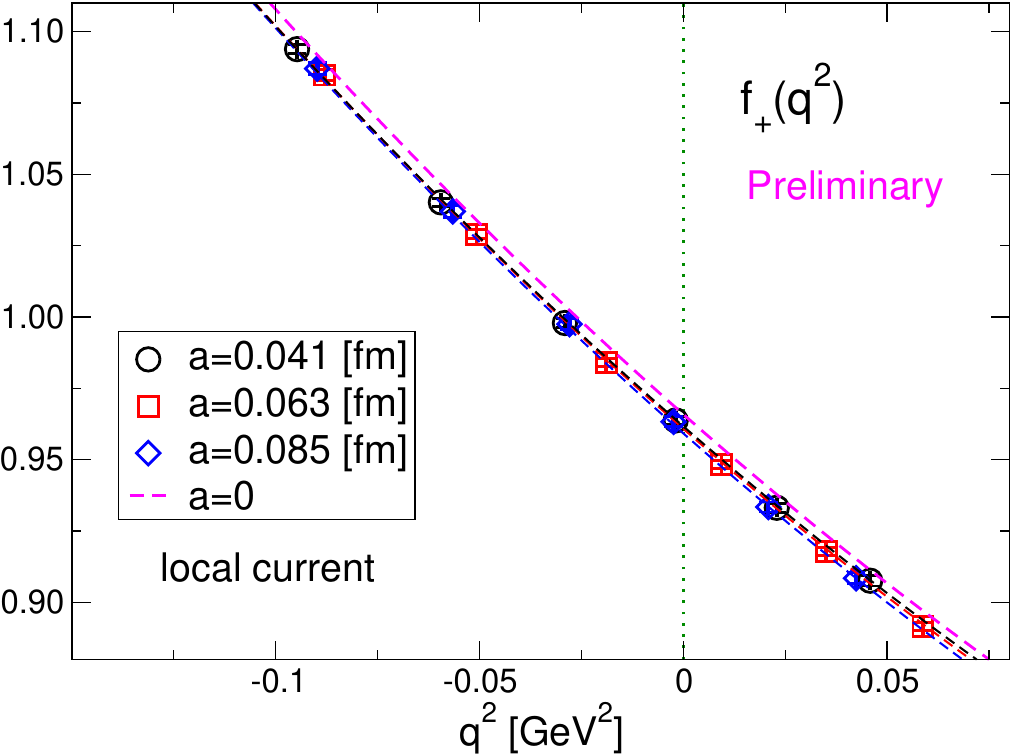}
 \hspace{2mm}
 \includegraphics*[scale=0.41]{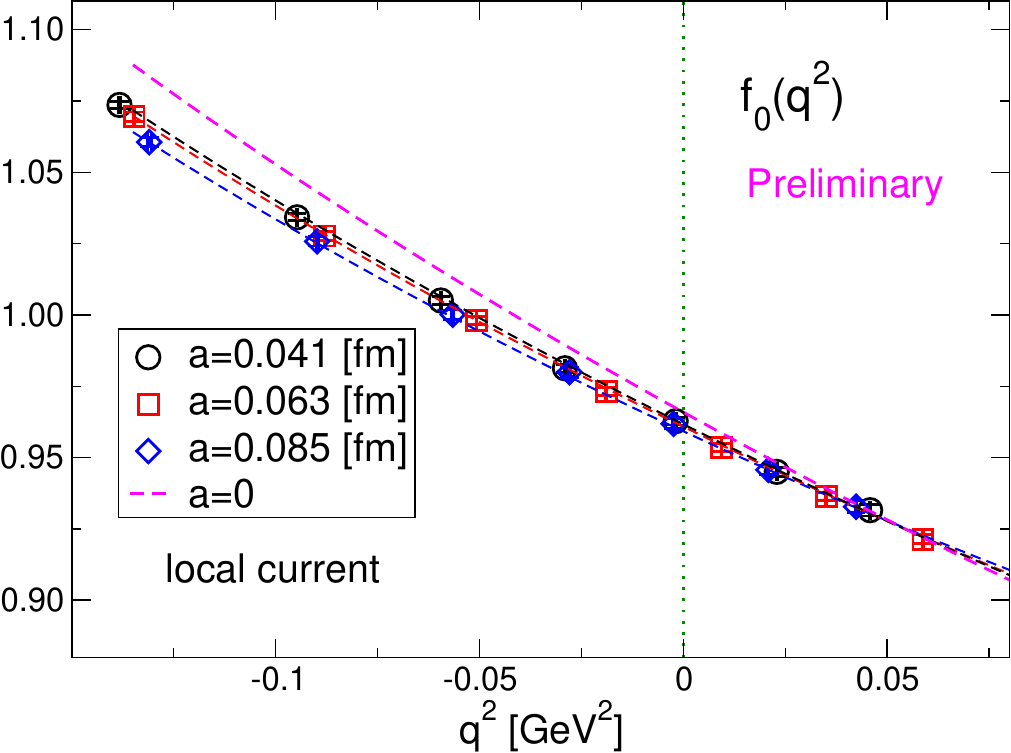}
 \caption{
Lattice spacing dependences for $f_+(q^2)$ (left) and $f_0(q^2)$ (right)
as a function of $q^2$ with the renormalized local vector current.
The different symbols represent data at the different lattice spacings.
The data at $a = 0.041$ fm are preliminary, while the other two data
are presented in Ref.~\cite{Ishikawa:2022ulx}.
The dashed curves express the result of the simultaneous fit described 
in the text.
The magenta curves correspond to the results in the continuum limit 
at the physical point.
  \label{fig:fpz_adep-loc}
 }
\end{figure}

Figure~\ref{fig:fpz_adep-loc} shows
our preliminary results for 
the $K_{\ell 3}$ form factors, $f_+(q^2)$ and $f_0(q^2)$,
at $a = 0.041$ fm including the previous results in Ref.~\cite{Ishikawa:2022ulx}
using the renormalized local vector current.
The data for $f_+(q^2)$ and $f_0(q^2)$ are measured precisely
at all the lattice spacings.
Thanks to the huge volume of the PACS10 configurations,
we can obtain several data near $q^2 = 0$ even using
the periodic boundary condition in the spatial directions.
The data with the conserved vector current have a similar quality
to those for the local vector current.

To determine the value of $f_+(0)$, 
which is important to obtain $|V_{us}|$ through the $K_{\ell 3}$ decay,
the data of the form factors are interpolated to $q^2 = 0$ and
also extrapolated to $a = 0$ simultaneously.
A fit form is based on the formulas at the next-to-leading order
in the chiral perturbation theory~\cite{Gasser:1984ux,Gasser:1984gg}, 
and we add correction terms
given by $(m_K^2 - m_\pi^2)^2$, $q^2$, and lattice spacing effects.
In the simultaneous fit, we use the data for $f_+(q^2)$ and $f_0(q^2)$ 
with both the local and conserved vector currents at all three
lattice spacings.
Figure~\ref{fig:fpz_adep-loc} presents
that the simultaneous fit works well in our local vector current data.
The fit results at each lattice spacing are denoted by
the curves in the same colors as the symbols.
Since the simulated $m_\pi$ and $m_K$ are different from the physical ones,
$m_{\pi^-} = 139.57061$ MeV and $m_{K^0} = 497.611$ MeV,
as shown in Table~\ref{tab:sim_param},
a short chiral extrapolation is carried out using the same formulas
as in the simultaneous fit.
The continuum limit results for $f_+(q^2)$ and $f_0(q^2)$
at the physical $m_\pi$ and $m_K$
are denoted by the magenta curves in the figure.

\subsection{Continuum extrapolation of $f_+(0)$}

The continuum extrapolation of $f_+(0)$ is discussed in this subsection,
which corresponds to the simultaneous fit result at $q^2 = 0$ 
explained in the previous subsection.

\begin{figure}[!t]
 \centering
 \includegraphics*[scale=0.41]{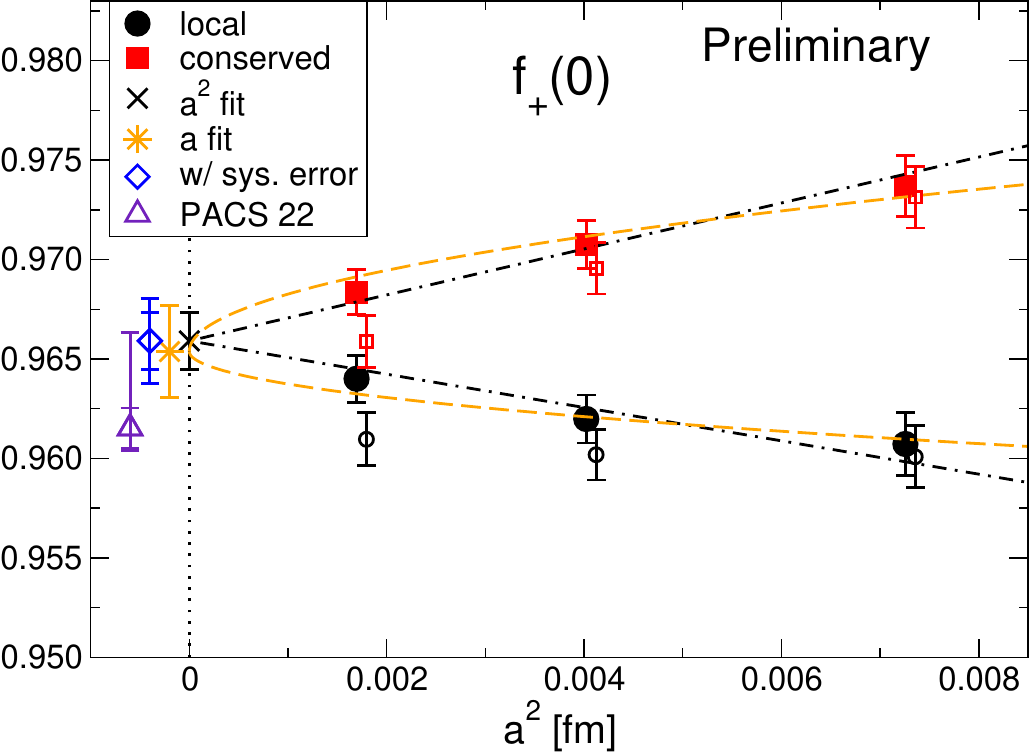}
 \hspace{2mm}
 \includegraphics*[scale=0.41]{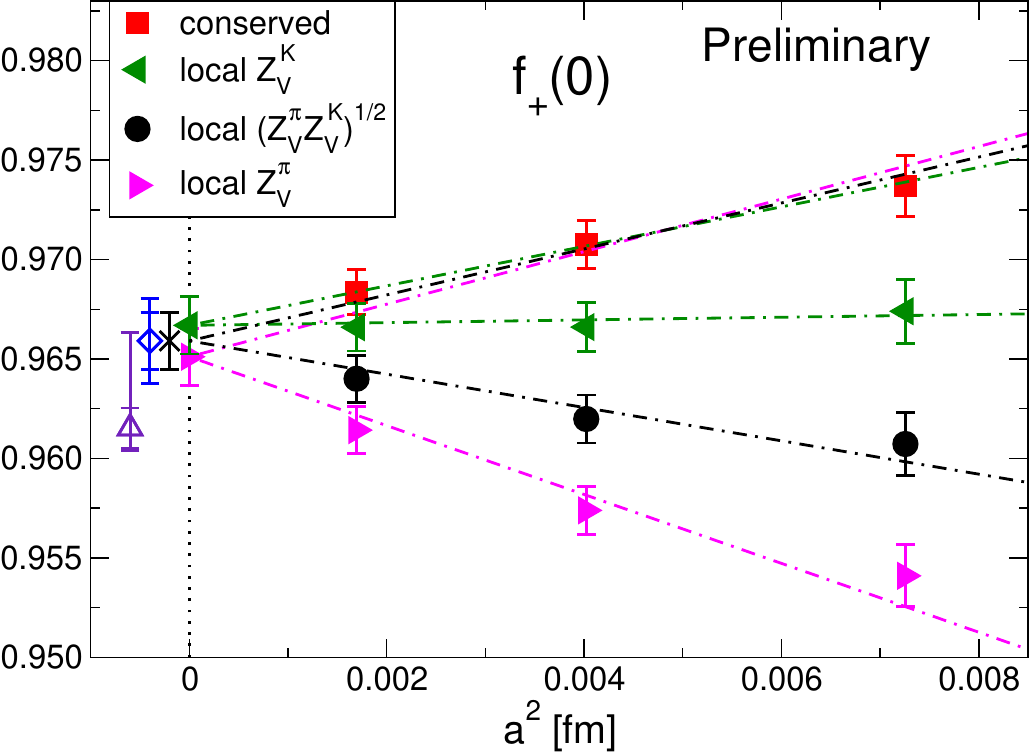}
 \caption{
{\bf Left :}
Preliminary results of $f_+(0)$ at the physical point
with the local (conserved) vector current denoted by black circle 
(red square) symbol as a function of $a$.
The open circle and square symbols represent the results at
the simulated meson masses.
The dashed-dot and dashed curves express continuum extrapolations
with quadratic and linear functions of $a$, respectively.
The blue diamond and violet up triangle symbols express
our preliminary and previous~\cite{Ishikawa:2022ulx} results. 
The inner and outer errors express the statistical and total errors.
The total error is evaluated by adding the statistical and systematic 
errors in quadrature.\\
{\bf Right :}
The same figure as the left one, but includes
the local vector current data with $Z_V^\pi$ and $Z_V^K$ denoted by
magenta right and green left triangle symbols, respectively.
The corresponding curves represent continuum extrapolations
for those data with a quadratic function of $a$.
  \label{fig:f+0_a0}
 }
\end{figure}

Before discussing the continuum extrapolation, we present
the effect of the chiral extrapolation in $f_+(0)$.
The effect of the short chiral extrapolation for $f_+(0)$ is shown in 
the left panel of Fig.~\ref{fig:f+0_a0}.
At the three lattice spacing, the open circle and square symbols
express the data at the simulated point
with the local and conserved currents, respectively.
The closed ones correspond to those at the physical point.
It is noted that those data are estimated from $q^2$ interpolations 
at each lattice spacing for discussing the chiral and continuum extrapolations.
The largest effect of the chiral extrapolation appears
at the smallest lattice spacing.
It is mainly caused by $m_K$, which differs from $m_{K^0}$
by about 3\% at the lattice spacing.

A continuum extrapolation of $f_+(0)$ in the simultaneous fit
is expressed by black dashed-dot curves in the left panel of 
Fig.~\ref{fig:f+0_a0}.
In the fit form, the finite lattice spacing effect is expressed by a quadratic
function of $a$,
and we constrain that the two data with the local and conserved currents should
coincide in the continuum limit.
We also perform a fit with a linear function of $a$ for 
the continuum extrapolation presented by orange dashed curves in the figure.
The result is consistent with that from the $a^2$ extrapolation.

Various different analyses are performed to
estimate systematic errors of $f_+(0)$.
An example of the analyses is using a different renormalization
factor in the local vector current data.
As explained in the last section, the renormalization factor
$Z_V$ is determined from a combination of $Z_V^\pi$ and $Z_V^K$
in eq.~(\ref{eq:zv}).
On the other hand, the renormalization factor can be determined from
each $Z_V^H$.
In a different analysis, the data with the local vector current
are replaced by the same data but using a different renormalization
factor, $Z_V^\pi$ or $Z_V^K$.
Those data are presented in the right panel of Fig.~\ref{fig:f+0_a0}
with the magenta right and green left triangle symbols.
The results in the continuum limit are well consistent with
the black cross symbol which is the same one as in the left panel.
Using results obtained from such different analyses,
the systematic error of $f_+(0)$ is estimated from
the maximum difference of the central value of the black cross
result from those in the different analyses.
The result with the total error including the systematic one 
is denoted by the blue diamond symbol in both the panels.
The result reasonably agrees with our previous result 
in Ref.~\cite{Ishikawa:2022ulx}
denoted by the violet up triangle symbol in the panels.

\begin{figure}[!t]
 \centering
 \includegraphics*[scale=0.40]{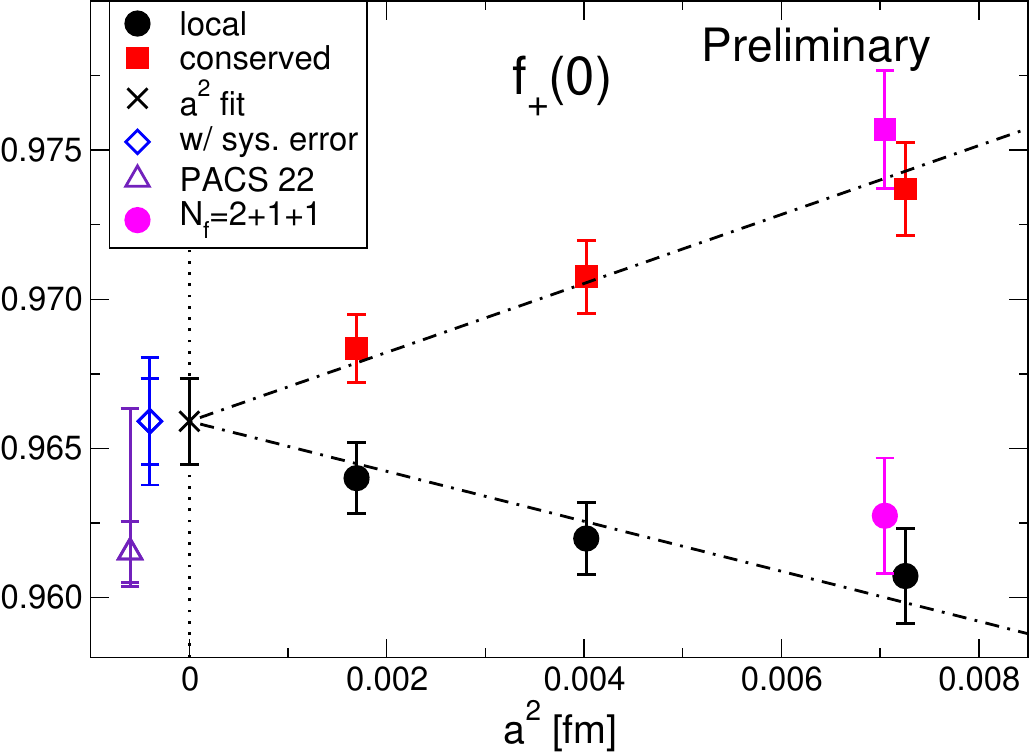}
 \caption{
The same figure as Fig.~\ref{fig:f+0_a0}, but contains 
the preliminary result from the PACS10$_c$ configuration
represented by magenta symbols.
  \label{fig:f00_PACS10c}
 }
\end{figure}

For a more precise evaluation of $f_+(0)$,
it is important to include the dynamical charm quark effect.
We have started to generate $N_f = 2+1+1$ PACS10 configurations,
named PACS10$_c$ configurations,
which satisfy the same condition as the PACS10 configurations but 
contain the dynamical charm quark effect.
So far only a preliminary result at the largest lattice spacing
has been obtained.
Figure~\ref{fig:f00_PACS10c} presents
a reasonable agreement of the PACS10$_c$ results with those 
from the $N_f = 2+1$ PACS10 configuration 
at a similar lattice spacing.
While we will not discuss the result further in this report,
we continue the calculation with PACS10$_c$ configurations
at smaller lattice spacings.

\begin{figure}[!t]
 \centering
 \includegraphics*[scale=0.40]{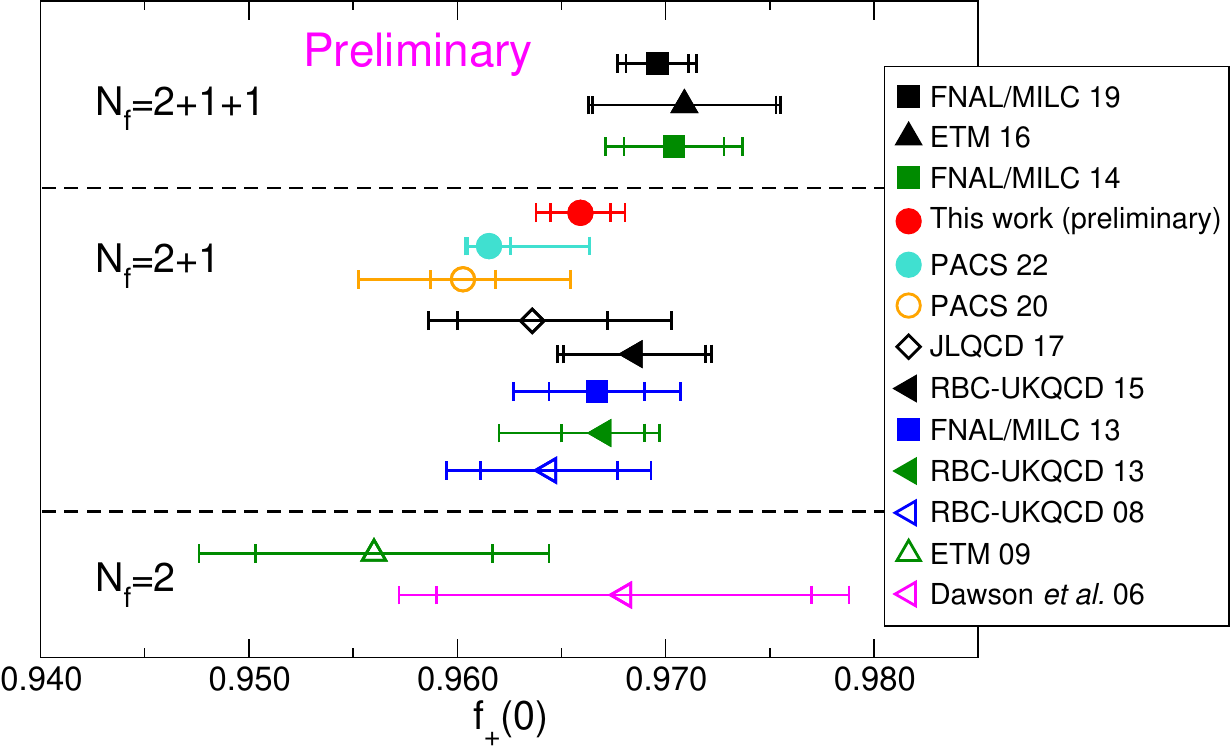}
 \caption{
Comparison of our preliminary result of $f_+(0)$ with
the previous results~\cite{Dawson:2006qc,Boyle:2007qe,Lubicz:2009ht,Bazavov:2012cd,Boyle:2013gsa,Bazavov:2013maa,Boyle:2015hfa,Carrasco:2016kpy,Aoki:2017spo,Bazavov:2018kjg,PACS:2019hxd,Ishikawa:2022ulx}.
The inner and outer errors
express the statistical and total errors.
The total error is evaluated by adding the statistical and systematic errors in quadrature.
The closed (open) symbols represent results 
in the continuum limit (at a finite lattice spacing).
  \label{fig:comp_f00}
 }
\end{figure}

In Fig.~\ref{fig:comp_f00},
our preliminary result of $f_+(0)$ is compared with
previous lattice calculations~\cite{Dawson:2006qc,Boyle:2007qe,Lubicz:2009ht,Bazavov:2012cd,Boyle:2013gsa,Bazavov:2013maa,Boyle:2015hfa,Carrasco:2016kpy,Aoki:2017spo,Bazavov:2018kjg}
including our previous results~\cite{PACS:2019hxd,Ishikawa:2022ulx}.
The preliminary result in this work is consistent with 
those in our previous calculations, and also
agrees with other lattice results within 2 $\sigma$.

\subsection{$|V_{us}|$ in the continuum limit}

\begin{figure}[!t]
 \centering
 \includegraphics*[scale=0.40]{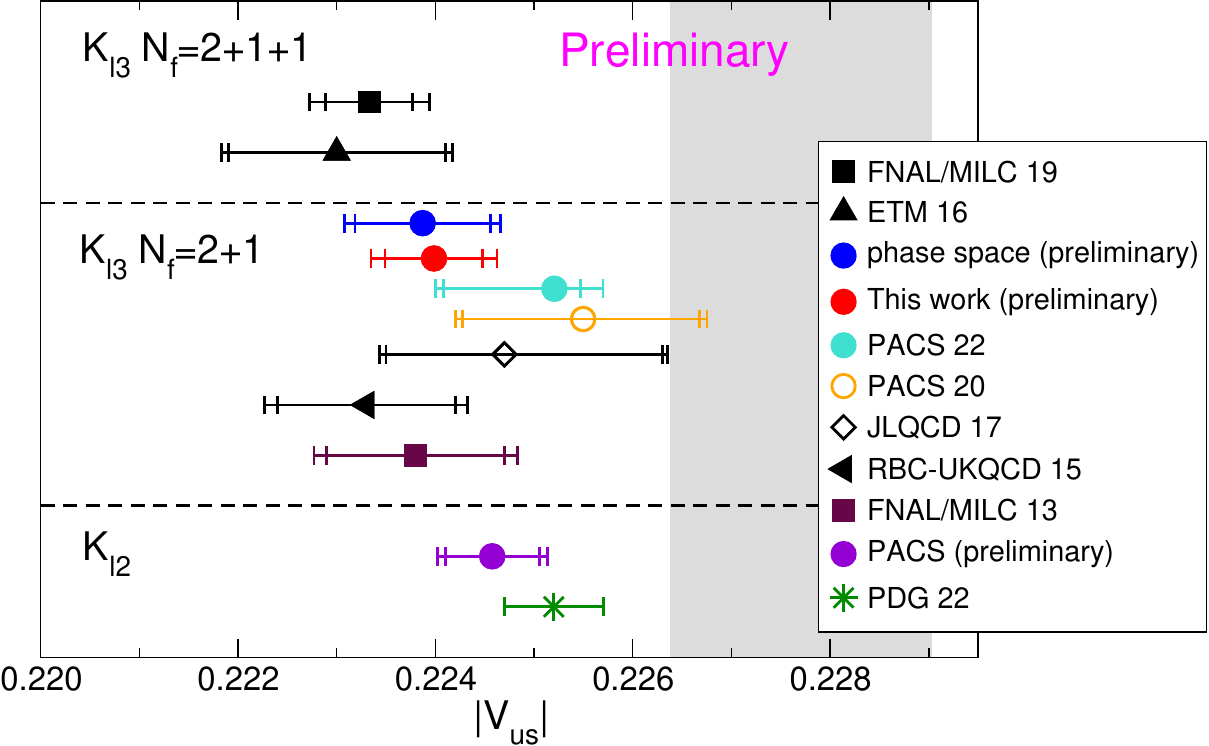}
 \caption{
Comparison of $|V_{us}|$ using our preliminary result of $f_+(0)$ with 
the previous results~\cite{Boyle:2013gsa,Boyle:2015hfa,Aoki:2017spo,Carrasco:2016kpy,Bazavov:2018kjg,PACS:2019hxd,Ishikawa:2022ulx}.
$|V_{us}|$ determined from the $K_{\ell 2}$ decay are also
plotted using $F_K/F_\pi$ of our preliminary result and 
PDG22~\cite{ParticleDataGroup:2022pth} together with
the one obtained from the phase space integral with our results
for $f_+(q^2)$ and $f_0(q^2)$.
The closed (open) symbols represent results 
in the continuum limit (at a finite lattice spacing).
The inner and outer errors
express the lattice QCD and total errors.
The total error is evaluated by adding the errors in the lattice QCD and experiment in quadrature.
The value of $|V_{us}|$ determined from the unitarity of the CKM matrix using $|V_{ud}|$ in Ref.~\cite{Hardy:2020qwl} is presented by the grey band.
  \label{fig:comp_vus}
 }
\end{figure}

The value of $|V_{us}|$ is determined using the result of $f_+(0)$
discussed in the previous subsection with
the experimental value $|V_{us}|f_+(0) = 0.21635(39)$~\cite{Seng:2021nar}.
The obtained $|V_{us}|$ using our preliminary result
is plotted in Fig.~\ref{fig:comp_vus}
together with the values using the previous lattice results~\cite{Boyle:2013gsa,Boyle:2015hfa,Aoki:2017spo,Carrasco:2016kpy,Bazavov:2018kjg,PACS:2019hxd,Ishikawa:2022ulx}.
Similar to the results of $f_+(0)$ in Fig.~\ref{fig:comp_f00}, 
our result is reasonably consistent with
the other results through the $K_{\ell 3}$ determination.

Our result of $|V_{us}|$ from $f_+(0)$ also reasonably agrees with
$|V_{us}|$ through the $K_{\ell 2}$ decay.
In this determination, $|V_{us}|$ is obtained 
from the ratio of the decay constants $F_K/F_\pi$ and the experimental value
$|V_{us}|F_K/|V_{ud}|F_\pi = 0.27683(35)$~\cite{DiCarlo:2019thl}.
In Fig.~\ref{fig:comp_vus}, we plot $|V_{us}|$ with our preliminary value
of $F_K/F_\pi$ obtained from the $N_f = 2+1$ PACS10 configurations,
and the one with $F_K/F_\pi$ in PDG22~\cite{ParticleDataGroup:2022pth}.
Furthermore, $|V_{us}|$ can be determined from 
the phase space integral~\cite{Leutwyler:1984je}
using our preliminary result of $f_+(q^2)$ and $f_0(q^2)$
in the continuum limit at the physical point.
The values of $|V_{us}|$ obtained from the $K_{\ell 2}$ determination
and phase space integral
are consistent with that from $f_+(0)$ as presented in Fig.~\ref{fig:comp_vus}.
On the other hand, $|V_{us}|$ estimated from the unitarity of the CKM
matrix using the value of $|V_{ud}|$ in Ref.~\cite{Hardy:2020qwl}
is about 2 $\sigma$ away from our $K_{\ell 3}$ determination,
which is shown by the grey band in the figure.

\section{Summary}

We have updated our calculation of the $K_{\ell 3}$ form factors 
using the three ensembles of the PACS10 configurations, 
which have more than (10 fm)$^4$ volumes very close to the physical point.
The finite lattice spacing effect of $f_+(0)$ is examined using the data
obtained from the local and conserved vector currents,
and also using the different renormalization factors for the local vector
current.
The systematic error of our preliminary result of $f_+(0)$ is estimated
from results with various different analyses.
We have observed a reasonable agreement of our preliminary result of $f_+(0)$
with other lattice calculations.
The value of $|V_{us}|$ determined with our $f_+(0)$
is reasonably consistent with those using other lattice results of $f_+(0)$,
the $K_{\ell 2}$ decay determination, and from the phase space integral
with our $q^2$ dependent form factors.
In contrast to them, $|V_{us}|$ estimated from the CKM unitarity
is different by about 2 $\sigma$ from our preliminary result.

For a more precise determination of $|V_{us}|$, an important future work
is the inclusion of the dynamical charm quark effect.
Toward this task, we have started to generate 
the $N_f=2+1+1$ PACS10$_c$ configurations.
We have obtained a consistent result from the PACS10$_c$ configuration 
with the PACS10 result at the largest lattice spacing.
We will calculate the $K_{\ell 3}$ form factors at smaller lattice spacings
with PACS10$_c$ configurations.

\acknowledgments{
Numerical calculations in this work were performed on Oakforest-PACS
in Joint Center for Advanced High Performance Computing (JCAHPC)
under Multidisciplinary Cooperative Research Program of Center for Computational Sciences, University of Tsukuba.
This research also used computational resources of Oakforest-PACS
by Information Technology Center of the University of Tokyo,
the Type II subsystem on supercomputer "Flow" at Information Technology Center, Nagoya University,
and of Fugaku by RIKEN CCS
through the HPCI System Research Project (Project ID: hp170022, hp180051, hp180072, hp180126, hp190025, hp190081, hp200062, hp200167, hp210112, hp220079, hp230199, hp240207).
The calculation employed OpenQCD system\footnote{http://luscher.web.cern.ch/luscher/openQCD/}.
This work was supported in part by Grants-in-Aid 
for Scientific Research from the Ministry of Education, Culture, Sports, 
Science and Technology (Nos. 19H01892, 23H01195, 23K25891) and
MEXT as ``Program for Promoting Researches on the Supercomputer Fugaku'' (Search for physics beyond the standard model using large-scale lattice QCD simulation and development of AI technology toward next-generation lattice QCD; Grant Number JPMXP1020230409).
This work was supported by the JLDG constructed over the SINET5 of NII.
}

\bibliographystyle{JHEP}
\bibliography{reference}

\providecommand{\href}[2]{#2}\begingroup\raggedright\begin{thebibliography}{10}

\bibitem{ParticleDataGroup:2022pth}
{\scshape Particle Data Group} collaboration, \emph{{Review of Particle
  Physics}}, \href{https://doi.org/10.1093/ptep/ptac097}{\emph{PTEP} {\bfseries
  2022} (2022) 083C01}.

\bibitem{Dawson:2006qc}
C.~Dawson, T.~Izubuchi, T.~Kaneko, S.~Sasaki and A.~Soni, \emph{{Vector form
  factor in K(l3) semileptonic decay with two flavors of dynamical domain-wall
  quarks}}, \href{https://doi.org/10.1103/PhysRevD.74.114502}{\emph{Phys. Rev.}
  {\bfseries D74} (2006) 114502}
  [\href{https://arxiv.org/abs/hep-ph/0607162}{{\ttfamily hep-ph/0607162}}].

\bibitem{Boyle:2007qe}
{\scshape RBC-UKQCD} collaboration, \emph{{K(l3) semileptonic form-factor from
  2+1 flavour lattice QCD}},
  \href{https://doi.org/10.1103/PhysRevLett.100.141601}{\emph{Phys. Rev. Lett.}
  {\bfseries 100} (2008) 141601}
  [\href{https://arxiv.org/abs/0710.5136}{{\ttfamily 0710.5136}}].

\bibitem{Lubicz:2009ht}
{\scshape ETM} collaboration, \emph{{$K \to \pi l \nu$ Semileptonic Form
  Factors from Two-Flavor Lattice QCD}},
  \href{https://doi.org/10.1103/PhysRevD.80.111502}{\emph{Phys. Rev.}
  {\bfseries D80} (2009) 111502}
  [\href{https://arxiv.org/abs/0906.4728}{{\ttfamily 0906.4728}}].

\bibitem{Bazavov:2012cd}
{\scshape Fermilab Lattice, MILC} collaboration, \emph{{Kaon semileptonic
  vector form factor and determination of $|V_{us}|$ using staggered
  fermions}}, \href{https://doi.org/10.1103/PhysRevD.87.073012}{\emph{Phys.
  Rev.} {\bfseries D87} (2013) 073012}
  [\href{https://arxiv.org/abs/1212.4993}{{\ttfamily 1212.4993}}].

\bibitem{Boyle:2013gsa}
{\scshape RBC-UKQCD} collaboration, \emph{{The kaon semileptonic form factor
  with near physical domain wall quarks}},
  \href{https://doi.org/10.1007/JHEP08(2013)132}{\emph{JHEP} {\bfseries 08}
  (2013) 132} [\href{https://arxiv.org/abs/1305.7217}{{\ttfamily 1305.7217}}].

\bibitem{Bazavov:2013maa}
{\scshape Fermilab Lattice, MILC} collaboration, \emph{{Determination of
  $|V_{us}|$ from a lattice-QCD calculation of the $K\to\pi\ell\nu$
  semileptonic form factor with physical quark masses}},
  \href{https://doi.org/10.1103/PhysRevLett.112.112001}{\emph{Phys. Rev. Lett.}
  {\bfseries 112} (2014) 112001}
  [\href{https://arxiv.org/abs/1312.1228}{{\ttfamily 1312.1228}}].

\bibitem{Boyle:2015hfa}
{\scshape RBC-UKQCD} collaboration, \emph{{The kaon semileptonic form factor in
  N$_{f}$ = 2 + 1 domain wall lattice QCD with physical light quark masses}},
  \href{https://doi.org/10.1007/JHEP06(2015)164}{\emph{JHEP} {\bfseries 06}
  (2015) 164} [\href{https://arxiv.org/abs/1504.01692}{{\ttfamily
  1504.01692}}].

\bibitem{Carrasco:2016kpy}
{\scshape ETM} collaboration, \emph{{$K \to \pi$ semileptonic form factors with
  $N_f=2+1+1$ twisted mass fermions}},
  \href{https://doi.org/10.1103/PhysRevD.93.114512}{\emph{Phys. Rev.}
  {\bfseries D93} (2016) 114512}
  [\href{https://arxiv.org/abs/1602.04113}{{\ttfamily 1602.04113}}].

\bibitem{Aoki:2017spo}
{\scshape JLQCD} collaboration, \emph{{Chiral behavior of $K \to \pi l \nu$
  decay form factors in lattice QCD with exact chiral symmetry}},
  \href{https://doi.org/10.1103/PhysRevD.96.034501}{\emph{Phys. Rev.}
  {\bfseries D96} (2017) 034501}
  [\href{https://arxiv.org/abs/1705.00884}{{\ttfamily 1705.00884}}].

\bibitem{Bazavov:2018kjg}
{\scshape Fermilab Lattice, MILC} collaboration, \emph{{$|V_{us}|$ from
  $K_{\ell 3}$ decay and four-flavor lattice QCD}},
  \href{https://doi.org/10.1103/PhysRevD.99.114509}{\emph{Phys. Rev.}
  {\bfseries D99} (2019) 114509}
  [\href{https://arxiv.org/abs/1809.02827}{{\ttfamily 1809.02827}}].

\bibitem{PACS:2019hxd}
{\scshape PACS} collaboration, \emph{{$K_{l3}$ form factors at the physical
  point on a $(10.9 fm)^3$ volume}},
  \href{https://doi.org/10.1103/PhysRevD.101.094504}{\emph{Phys. Rev. D}
  {\bfseries 101} (2020) 094504}
  [\href{https://arxiv.org/abs/1912.13127}{{\ttfamily 1912.13127}}].

\bibitem{Ishikawa:2022ulx}
{\scshape PACS} collaboration, \emph{{$K_{\ell 3}$ form factors at the physical
  point: Toward the continuum limit}},
  \href{https://doi.org/10.1103/PhysRevD.106.094501}{\emph{Phys. Rev. D}
  {\bfseries 106} (2022) 094501}
  [\href{https://arxiv.org/abs/2206.08654}{{\ttfamily 2206.08654}}].

\bibitem{Yamazaki:2023swq}
{\scshape PACS} collaboration, \emph{{$|V_{us}|$ from kaon semileptonic form
  factor in $N_f = 2+1$ QCD at the physical point on (10 fm)$^4$}},
  \href{https://doi.org/10.22323/1.453.0276}{\emph{PoS} {\bfseries LATTICE2023}
  (2024) 276} [\href{https://arxiv.org/abs/2311.16755}{{\ttfamily
  2311.16755}}].

\bibitem{Iwasaki:2011jk}
Y.~Iwasaki, \emph{{Renormalization Group Analysis of Lattice Theories and
  Improved Lattice Action. II -- four-dimensional non-abelian SU(N) gauge
  model}},  \href{https://arxiv.org/abs/1111.7054}{{\ttfamily 1111.7054}}.

\bibitem{Morningstar:2003gk}
C.~Morningstar and M.J.~Peardon, \emph{{Analytic smearing of SU(3) link
  variables in lattice QCD}},
  \href{https://doi.org/10.1103/PhysRevD.69.054501}{\emph{Phys. Rev.}
  {\bfseries D69} (2004) 054501}
  [\href{https://arxiv.org/abs/hep-lat/0311018}{{\ttfamily hep-lat/0311018}}].

\bibitem{Boyle:2008yd}
{\scshape RBC-UKQCD} collaboration, \emph{{The Pion's electromagnetic
  form-factor at small momentum transfer in full lattice QCD}},
  \href{https://doi.org/10.1088/1126-6708/2008/07/112}{\emph{JHEP} {\bfseries
  07} (2008) 112} [\href{https://arxiv.org/abs/0804.3971}{{\ttfamily
  0804.3971}}].

\bibitem{Kakazu:2017fhv}
{\scshape PACS} collaboration, \emph{{Electromagnetic pion form factor near
  physical point in $N_f=2+1$ lattice QCD}},
  \href{https://doi.org/10.22323/1.256.0160}{\emph{PoS} {\bfseries LATTICE2016}
  (2017) 160}.

\bibitem{Gasser:1984ux}
J.~Gasser and H.~Leutwyler, \emph{{Low-Energy Expansion of Meson
  Form-Factors}},
  \href{https://doi.org/10.1016/0550-3213(85)90493-6}{\emph{Nucl. Phys.}
  {\bfseries B250} (1985) 517}.

\bibitem{Gasser:1984gg}
J.~Gasser and H.~Leutwyler, \emph{{Chiral Perturbation Theory: Expansions in
  the Mass of the Strange Quark}},
  \href{https://doi.org/10.1016/0550-3213(85)90492-4}{\emph{Nucl. Phys.}
  {\bfseries B250} (1985) 465}.

\bibitem{Seng:2021nar}
C.-Y.~Seng, D.~Galviz, W.J.~Marciano and U.-G.~Mei\ss{}ner, \emph{{Update on
  |Vus| and |Vus/Vud| from semileptonic kaon and pion decays}},
  \href{https://doi.org/10.1103/PhysRevD.105.013005}{\emph{Phys. Rev. D}
  {\bfseries 105} (2022) 013005}
  [\href{https://arxiv.org/abs/2107.14708}{{\ttfamily 2107.14708}}].

\bibitem{DiCarlo:2019thl}
M.~Di~Carlo, D.~Giusti, V.~Lubicz, G.~Martinelli, C.T.~Sachrajda, F.~Sanfilippo
  et~al., \emph{{Light-meson leptonic decay rates in lattice QCD+QED}},
  \href{https://doi.org/10.1103/PhysRevD.100.034514}{\emph{Phys. Rev.}
  {\bfseries D100} (2019) 034514}
  [\href{https://arxiv.org/abs/1904.08731}{{\ttfamily 1904.08731}}].

\bibitem{Leutwyler:1984je}
H.~Leutwyler and M.~Roos, \emph{{Determination of the Elements V(us) and V(ud)
  of the Kobayashi-Maskawa Matrix}},
  \href{https://doi.org/10.1007/BF01571961}{\emph{Z. Phys.} {\bfseries C25}
  (1984) 91}.

\bibitem{Hardy:2020qwl}
J.C.~Hardy and I.S.~Towner, \emph{{Superallowed $0^+ \to 0^+$ nuclear $\beta$
  decays: 2020 critical survey, with implications for V$_{ud}$ and CKM
  unitarity}}, \href{https://doi.org/10.1103/PhysRevC.102.045501}{\emph{Phys.
  Rev. C} {\bfseries 102} (2020) 045501}.

\end{thebibliography}\endgroup

\end{document}